# The evolution-dominated hydrodynamic model and the pseudorapidity distributions in high energy physics

Z. J. Jiang[*], H. L. Zhang, J. Wang and K. Ma

College of Science, University of Shanghai for Science and Technology, Shanghai 200093, China

[*]E-mail: jzj265@163.com

**Abstract**

By taking into account the effects of leading particles, we discuss the pseudorapidity distributions of the charged particles produced in high energy heavy ion collisions in the context of evolution-dominated hydrodynamic model. The leading particles are supposed to have a Gaussian rapidity distribution normalized to the number of participants. A comparison is made between the theoretical results and the experimental measurements performed by BRAHMS and PHOBOS Collaboration at BNL-RHIC in Au-Au and Cu-Cu collisions at $\sqrt{s_{NN}}$ =200 GeV and by ALICE Collaboration at CERN-LHC in Pb-Pb collisions at $\sqrt{s_{NN}}$ =2.76 TeV.

PACS number(s): 25.75.Ag, 25.75.Ld, 24.10.Nz

## 1. Introduction

Along with the successful description of elliptic flow and multiplicity production in heavy ion collisions [1-4], relativistic hydrodynamics has now been widely accepted as one of the most important tools for understanding the space-time evolution of the matter created in collisions [5-11]. With the specified initial conditions, the equation of state and the freeze-out conditions, the motion of fluid relies only on the local energy-momentum conservation and the assumption of local thermal equilibrium. From this point of view, hydrodynamics is simple and powerful. However, on the other hand, the initial conditions, the equation of state and the freeze-out conditions of fluid are not well known. Worse still is that the partial differential equations of relativistic hydrodynamics are highly non-linear and coupled. It is a very hard thing to solve them analytically. From this point of view, hydrodynamics is tremendously complicated. This is the reason why the progress in finding exact hydrodynamic solutions is not going well. Up till now, most part of this work only involves in 1+1 dimensional flows for the perfect fluid with the simple equation of state. The 3+1 dimensional hydrodynamics is less developed, and no general exact solutions are known so far.

The first exact solution of 1+1 dimensional expansions is given by L. D. Landau, S. Z. Belenkij and I. M. Khalatnikov about 60 years ago [12-14]. This solution is for an accelerated perfect fluid being initially at rest. The obtained result is a very unpleasant one since it is presented in a rather implicit way. However, the good news is that, from this complicated solution, Landau managed to extract approximately the rapidity distribution of charged particles, which is in generally accordance with the observations made at BNL-RHIC [15-17].

The first exact analytical solution of 1+1 dimensional hydrodynamics is given by R. C. Hwa and J. D. Bjorken [18, 19]. This solution is for an accelerationless system with Lorentz invariant initial condition. The result got in this way is simple and explicit. From this solution, Bjorken was able to get a simple estimate of the initial energy density reached in collisions from the final hadronic observables. This makes the energy density be measurable in experiment. However, the sad news is that the invariant rapidity distribution resulted from this model is at variance with experimental observations.

Among other theoretical models, it is worth to notice that a series of rigorous solutions of relativistic hydrodynamics has been found in recent years. By utilizing the scheme of the Khalatnikov potential, Ref. [20] solved the evolution-dominated hydrodynamics exactly, and obtained a set of analytical solutions for a perfect fluid with linear equation of state. Refs. [21-23] used the variational method to generalize the Hwa-Bjorken model for accelerationless system to the one for accelerated system, and obtained a new class of exact analytical solutions of relativistic hydrodynamics. By generalizing Hwa-Bjorken in-out ansatz for liquid trajectories, Ref. [24] also presented a family of exact solutions under different freeze-out conditions. By using scalar field theory, Ref. [25] gave a set of exact solutions describing non-stationary and inhomogeneous flows of the perfect fluid under different linear and non-linear equations of state. By applying the lattice QCD equation of state, Ref. [26] generated a group of exact solutions for accelerationless system with ellipsoidal symmetry.

In the present paper, by using the evolution-dominated hydrodynamics [20] and taking into account the contribution from leading particles, we shall discuss the pseudorapidity distributions of the charged particles produced in heavy ion collisions. We shall first give a brief introduction to the evolution-dominated hydrodynamics in section 2. Then, in section 3, a comparison is made between the theoretical results and experiment measurements carried out by BRAHMS and PHOBOS Collaboration at BNL-RHIC in Au-Au and Cu-Cu collisions at $\sqrt{s_{NN}}$ =200 GeV [15-17, 27] and by ALICE Collaboration at CERN-LHC in Pb-Pb collisions at $\sqrt{s_{NN}}$ =2.76 TeV [28]. This comparison indicates that, in addition to particles resulted from fluid evolution, leading particles are also essential in describing experimental data. Only after both effects are taken into account together, can the experimental observations be described properly. The last section is traditionally about conclusions.

## 2. A brief introduction to evolution-dominated hydrodynamics

Here, for the purpose of completion and applications, we shall list out the key ingredients of the evolution-dominated hydrodynamics [20].

(1) The 1+1 expansion of a perfect fluid obeys equation

$$\begin{cases} \dfrac{e^{2y}-1}{2}\dfrac{\partial(\varepsilon+p)}{\partial z^+} + e^{2y}(\varepsilon+p)\dfrac{\partial y}{\partial z^+} + \dfrac{1-e^{-2y}}{2}\dfrac{\partial(\varepsilon+p)}{\partial z^-} \\ \qquad\qquad\qquad + e^{-2y}(\varepsilon+p)\dfrac{\partial y}{\partial z^-} + \dfrac{\partial p}{\partial z^+} - \dfrac{\partial p}{\partial z^-} = 0, \\ \dfrac{e^{2y}+1}{2}\dfrac{\partial(\varepsilon+p)}{\partial z^+} + e^{2y}(\varepsilon+p)\dfrac{\partial y}{\partial z^+} + \dfrac{1+e^{-2y}}{2}\dfrac{\partial(\varepsilon+p)}{\partial z^-} \\ \qquad\qquad\qquad - e^{-2y}(\varepsilon+p)\dfrac{\partial y}{\partial z^-} - \dfrac{\partial p}{\partial z^+} - \dfrac{\partial p}{\partial z^-} = 0, \end{cases} \quad (1)$$

where $\varepsilon$, $p$ and $y$ are respectively the energy density, pressure and ordinary rapidity of fluid. $z^{\pm}=t\pm z=x^0\pm x^1=\tau e^{\pm\eta}$ is the light-cone coordinates, $\tau=\sqrt{z^+z^-}$ is the proper time, and $\eta=1/2\ln\left(z^+/z^-\right)$ is the space-time rapidity of fluid.

Eq. (1) is a complicated, non-linear and coupled one. In order to solve it, one introduces Khalatnikov potential $\chi$, which relates to $z^{\pm}$, $\tau$ and $\eta$ by equations

$$\begin{cases} z^{\pm}(\theta,y) = \dfrac{1}{2T_0} e^{\theta \pm y} \left( \dfrac{\partial \chi}{\partial \theta} \pm \dfrac{\partial \chi}{\partial y} \right), \\ \tau(\theta,y) = \dfrac{e^{\theta}}{2T_0} \sqrt{\left(\dfrac{\partial \chi}{\partial \theta}\right)^2 - \left(\dfrac{\partial \chi}{\partial y}\right)^2}, \\ \eta(\theta,y) = y + \dfrac{1}{2} \ln\left( \dfrac{\partial \chi/\partial \theta + \partial \chi/\partial y}{\partial \chi/\partial \theta - \partial \chi/\partial y} \right), \end{cases} \quad (2)$$

where

$$\theta = \ln\left(\frac{T_0}{T}\right), \quad (3)$$

$T$ is the temperature of liquid, and $T_0$ is its initial scale. In terms of $\chi$, Eq. (1) can be reduced to

$$\frac{\partial^2 \chi(\theta,y)}{\partial \theta^2} - \left[g(\theta) - 1\right] \frac{\partial \chi(\theta,y)}{\partial \theta} - g(\theta) \frac{\partial^2 \chi(\theta,y)}{\partial y^2} = 0, \quad (4)$$

where $1/\sqrt{g(\theta)} = c_s(\theta)$ is the speed of sound. The above equation is now a linear second-order partial differential equation, which works for any form of $g(\theta)$.

(2) In case of $g(\theta) = g = $ constant, the solution of Eq. (4) takes the form as

$$\chi(\theta,y) = \frac{e^{\frac{g-1}{2}\theta}}{4\sqrt{g}} \int dy' \int_0^{\theta - (y-y')/\sqrt{g}} d\theta' F(\theta',y') I_0\left( \frac{g-1}{2} \sqrt{(\theta - \theta')^2 - \frac{(y-y')^2}{g}} \right), \quad (5)$$

where $F(\theta',y')$ stands for the distribution of sources of hydrodynamic flow.

Experimental investigations have shown that the speed of sound is a constant of about $c_s = 0.35$ or $g = 8.16$, which is almost independent of interaction energy and system [29-32].

(3) For evolution-dominated hydrodynamics, the distribution of source takes the form as [14, 20, 33]

$$F(\theta',y') = Ce^{-\frac{g+1}{2}\theta'} \Theta(\theta') \delta(y'), \quad (6)$$

where $C$ is a constant. Inserting it into Eq. (5), it reads

$$\chi(\theta,y) = Ce^{-\theta} \int_{y/\sqrt{g}}^{\theta} d\theta' e^{\frac{g+1}{2}\theta'} I_0\left( \frac{g-1}{2} \sqrt{\theta'^2 - \frac{y^2}{g}} \right). \quad (7)$$

In heavy ion collisions at high energy, owing to the violent compression of collision system along beam direction, the initial pressure gradient of created matter in this direction is very large.

By contrast, the effect of initial flow of sources is negligible. The motion of liquid is mainly dominated by the following evolution. Hence, the evolution-dominated hydrodynamics should reflect the reality of expansion of fluid.

(4) The freeze-out of fluid is assumed to take place at a space-like hypersurface with a fixed temperature of $T_{FO}$. From this assumption together with the direct proportional relation between the number of charged particles and entropy, we can get the rapidity distribution of charged particles

$$\frac{dN_{Fluid}}{dy} \propto \left. \frac{\partial^2 \chi(\theta,y)}{\partial \theta^2} + \frac{\partial \chi(\theta,y)}{\partial \theta} \right|_{\theta=\theta_{FO}}, \tag{8}$$

where $\theta_{FO} = \ln(T_0/T_{FO})$, which is related to the initial temperature of fluid and is therefore dependent on the incident energy and collision centrality. Its specific value can be determined by comparing with experimental data.

Substituting Eq. (7) into above equation, it becomes

$$\frac{dN_{Fluid}(b,\sqrt{s_{NN}},y)}{dy} = C(b,\sqrt{s_{NN}}) \left[ I_0\left(\frac{g-1}{2}\sqrt{\theta_{FO}^2 - \frac{y^2}{g}}\right) + \frac{\theta_{FO}}{\sqrt{\theta_{FO}^2 - y^2/g}} I_1\left(\frac{g-1}{2}\sqrt{\theta_{FO}^2 - \frac{y^2}{g}}\right) \right], \tag{9}$$

where $C(b,\sqrt{s_{NN}})$, independent of rapidity $y$, is an overall normalization constant. $b$ is the impact parameter, and $\sqrt{s_{NN}}$ is the center-of-mass energy per pair of nucleons.

## 3. Comparison with experimental measurements and the rapidity distributions of leading particles

Figure 1 shows the rapidity distributions for $\pi^+$, $\pi^-$, $K^+$, $K^-$, $p$ and $\bar{p}$ produced in central Au-Au collisions at $\sqrt{s_{NN}}$ =200 GeV. The scattered symbols are the experimental data [15-17]. The solid curves are the theoretical results from Eq. (9). In calculations, the parameter $\theta_{FO}$ takes the value of $\theta_{FO}$ =2.23. We can see from this figure that, except for proton $p$, Eq. (9) fits well with experimental measurements. For proton $p$, experimental data show an evident uplift in the rapidity interval between $y = 2.0$ and $3.0$. This may be resulted from parts of leading particles, which is free from the description of hydrodynamics. Hence, in order to match up with experimental data, we should take these leading particles into account separately.

Considering that, for a given incident energy, the leading particles in each time of nucleus-nucleus collisions have approximately the same amount of energy, then, according to the central limit theorem [34, 35], the leading particles should follow the Gaussian rapidity distribution. That is

$$\frac{\mathrm{d}N_{\text{Lead}}\left(b,\sqrt{s_{\text{NN}}},y\right)}{\mathrm{d}y} = \frac{N_{\text{Lead}}\left(b,\sqrt{s_{\text{NN}}}\right)}{\sqrt{2\pi}\sigma}\exp\left\{-\frac{\left[|y|-y_0\left(b,\sqrt{s_{\text{NN}}}\right)\right]^2}{2\sigma^2}\right\}, \qquad (10)$$

where $y_0\left(b,\sqrt{s_{\text{NN}}}\right)$ and $\sigma$ are respectively the central position and width of distribution. In fact, as is known to all, the rapidity distribution of any charged particles produced in heavy ion collisions can be well represented by Gaussian form [15-17; also confer the shapes of the curves in Figure 1]. It is obvious that $y_0\left(b,\sqrt{s_{\text{NN}}}\right)$ should increase with incident energy and centrality cut. However, $\sigma$ should not apparently depend on them. This is due to the fact that the relative energy differences among leading particles should not be too much for different incident energies and centrality cuts. $N_{\text{Lead}}\left(b,\sqrt{s_{\text{NN}}}\right)$ in Eq. (10) is the number of leading particles. It is the function of energy and centrality. Inspired by nucleon-nucleon, such as $p$-$p$ collisions, where there are only two leading particles, the number of leading particles in nucleus-nucleus collisions should equal the number of participants, which can be evaluated by formula [36]

$$N_{\text{Part}}\left(b,\sqrt{s_{\text{NN}}}\right) = \int n_{\text{Part}}\left(b,\sqrt{s_{\text{NN}}},s\right)\mathrm{d}^2 s, \qquad (11)$$

where $s$ is the coordinates in the overlap region measured from the center of one nucleus.

$$n_{\text{Part}}\left(b,\sqrt{s_{\text{NN}}},s\right) = T_A(s)\left\{1-\exp\left[-\sigma_{\text{NN}}^{\text{in}}\left(\sqrt{s_{\text{NN}}}\right)T_B(s-b)\right]\right\} + T_B(s-b)\left\{1-\exp\left[-\sigma_{\text{NN}}^{\text{in}}\left(\sqrt{s_{\text{NN}}}\right)T_A(s)\right]\right\},$$

where $\sigma_{\text{NN}}^{\text{in}}\left(\sqrt{s_{\text{NN}}}\right)$ is the inelastic nucleon-nucleon cross section. It increases slowly with energy. Such as, for $\sqrt{s_{\text{NN}}}$ =200 GeV, $\sigma_{\text{NN}}^{\text{in}}$ =42 mb [37], and for $\sqrt{s_{\text{NN}}}$ =2.76 TeV, $\sigma_{\text{NN}}^{\text{in}} = 64 \pm 5$ mb [38]. The subscripts A and B in the above equation denote the projectile and target nucleus, respectively. $T(s)$ is the thickness function defined as

$$T(s) = \int \rho(s,z) \mathrm{d}z,$$

where

$$\rho(r) = \frac{\rho_0}{1 + \exp[(r - r_0)/a]}$$

is the Woods-Saxon distribution of nuclear density. $a$ and $r_0$ are respectively the skin depth and radius of nucleus. In this paper, they take the values of $a$=0.54 fm and $r_0 = 1.12 A^{1/3} - 0.86 A^{-1/3}$ fm [39], where $A$ is the mass number of nucleus.

Tables 1 and 2 show the mean numbers of total participants in different centrality Au-Au and Cu-Cu collisions at $\sqrt{s_{NN}}$ = 200 GeV and Pb-Pb collisions at $\sqrt{s_{NN}}$ =2.76 TeV. The numbers with and without errors are those given by experiments [27, 28] and Eq. (11), respectively. Due to space constraints, Table 1 only shows the numbers in the first nine centrality cuts. It can be seen that both sets of numbers coincide well.

Having the rapidity distributions of Eqs. (9) and (10), the pseudorapidity distribution measured in experiments can be expressed as [40]

$$\frac{\mathrm{d}N(b,\sqrt{s_{NN}},\eta)}{\mathrm{d}\eta} = \sqrt{1 - \frac{m^2}{m_T^2 \cosh^2 y}} \frac{\mathrm{d}N(b,\sqrt{s_{NN}},y)}{\mathrm{d}y}, \qquad (12)$$

$$y = \frac{1}{2} \ln \left[ \frac{\sqrt{p_T^2 \cosh^2 \eta + m^2} + p_T \sinh \eta}{\sqrt{p_T^2 \cosh^2 \eta + m^2} - p_T \sinh \eta} \right], \qquad (13)$$

where $p_T$ is the transverse momentum, $m_T = \sqrt{m^2 + p_T^2}$ is the transverse mass, and

$$\frac{\mathrm{d}N(b,\sqrt{s_{NN}},y)}{\mathrm{d}y} = \frac{\mathrm{d}N_{\mathrm{Fluid}}(b,\sqrt{s_{NN}},y)}{\mathrm{d}y} + \frac{\mathrm{d}N_{\mathrm{Lead}}(b,\sqrt{s_{NN}},y)}{\mathrm{d}y} \qquad (14)$$

is the total rapidity distribution from both fluid evolution and leading particles.

Substituting Eq. (14) into (12), we can get the pseudorapidity distributions of charged particles. Figures 2, 3 and 4 show such distributions in different centrality Au-Au and Cu-Cu collisions at $\sqrt{s_{NN}}$ =200 GeV and Pb-Pb collisions at 2.76 TeV, respectively. The solid dots in figures are the experimental measurements [27, 28]. The dashed curves are the results got from evolution-dominated hydrodynamics of Eq. (9). The dotted curves are the results obtained from leading particles of Eq. (10). The solid curves are the results achieved from Eq. (14), that is, the sums of dashed and dotted curves. It can be seen that the theoretical results are well consistent with experimental measurements.

In calculations, the parameter $\theta_{FO}$ in Eq. (9) takes the values of 2.80 in the first three centrality cuts, 2.98 in the following six ones and 3.17 in the last two ones in Au-Au collisions. In Cu-Cu collisions, $\theta_{FO}$ takes the value of 2.95 in the first three centrality cuts, 3.15 in the following six ones and 3.53 in the last three ones. In Pb-Pb collisions, $\theta_{FO}$ takes the value of 5.85 for the first two centrality cuts and 6.04 for the second two ones. It can be seen that $\theta_{FO}$ increases with incident energy and centrality cuts. The width parameter $\sigma$ in Eq. (10) takes a constant of 0.85 for all three kinds of collision systems in different centrality cuts. As the analyses given above, $\sigma$ is independent of incident energy and centrality cut. What is more, it is also independent of collision system. The center parameter $y_0$ in Eq. (10) takes the values as listed in Tables 1 and 2. As stated early, $y_0$ increases with energy and centrality cut. From Table 1, we can see that, for a given incident energy and centrality cut, $y_0$ decreases with the increasing of nucleus size. This can be understood if we notice the fact that the larger the nucleus size, the more collisions will the participants undergo. Hence, the final leading particles will lose more energy or have smaller $y_0$. The fitting value of $y_0 = 2.63$ in the top 3% most central Au-Au collisions is in accordance with the experimental observation shown in Fig. 1, which indicates that the leading particles are mainly in the range between $y = 2$ and 3. Experimental investigations also have shown that [41], in the top 5% most central Au-Au collisions at $\sqrt{s_{NN}} = 200\,\text{GeV}$, the rapidity loss of participants is up to $<\delta y> \approx 2.45$, then the leading particles should locate at

$$y_0 = y_{beam} - <\delta y> = 5.36 - 2.45 = 2.91.$$

Seeing that the smaller centrality cut considered in our analysis, our above fitting result is also consistent with this measurement. Comparing the fitting values of $y_0$ in Tables 1 and 2, we can see that $y_0$ increases very slowly with energy. This is in agreement with the experimental observations that the rapidity loss of participants seem to saturate from SPS energies.

## 4. Conclusions

The charged particles produced in heavy ion collisions are divided into two parts. One is from the hot and dense matter created in collisions. The other is from leading particles.

Compared with the effect of pressure gradient, the effect of initial flow of the hot and dense matter is negligible. The motion of this matter is mainly governed by the evolution of fluid. This thus guarantees the rationality of evolution-dominated hydrodynamics. With the scheme of Khalatnikov potential, this theoretical model can be solved exactly, and the rapidity distribution of charged particles can be expressed in a simple analytical form in terms of 0th and 1st order modified Bessel function of the first kind with only two parameters $g = 1/c_s^2$ and $\theta_{\text{FO}} = \ln(T_0/T_{\text{FO}})$. $g$ takes the value from experiments. $\theta_{\text{FO}}$ is fixed by fitting with experimental data.

For leading particles, we assume that the rapidity distribution of them possesses the Gaussian form with the normalization constant being equal to the number of participants, which can be figured out in theory. This assumption is based on the consideration that, for a given incident energy, the leading particles have about the same energy, and coincides with the fact that any kind of charged particles takes on well the Gaussian form of rapidity distribution. It is interested to notice that the width of Gaussian rapidity distribution $\sigma$ is irrelevant to the incident energy, centrality cut and collision system. The fitting values of $y_0$, the central positions of Gaussian rapidity distribution, are in good accordance with experimental data.

Comparing with the experimental measurements made by BRAHMS and PHOBOS Collaboration at BNL-RHIC in Au-Au and Cu-Cu collisions at $\sqrt{s_{\text{NN}}} = 200$ GeV and by ALICE Collaboration at CERN-LHC in Pb-Pb collisions at $\sqrt{s_{\text{NN}}} = 2.76$ TeV, we can see that the total contributions from both evolution-dominated hydrodynamics and leading particles are well consistent with experimental data.

## Acknowledgments

This work is partly supported by the Transformation Project of Science and Technology of Shanghai Baoshan District with Grant No. CXY-2012-25 and by the Shanghai Leading Academic

Discipline Project with Grant No. XTKX 2012.

## References

[1] J. Y. Ollitrault, *Phys*. *Rev*. D **46** 229 (1992).

[2] S. S. Adler *et al.* (PHENIX Collaboration), *Phys*. *Rev*. *Lett*. **91** 182301 (2003).

[3] K. Aamodt *et al.* (ALICE Collaboration), *Phys*. *Rev*. *Lett*. **107** 032301 (2011).

[4] P. Steinberg, *Nucl*. *Phys*. A **752** 423 (2005).

[5] A. Bialas and R. Peschanski, *Phys*. *Rev*. C **83** 054905 (2011).

[6] C. Y. Wong, *Phys. Rev. C* **78** 054902 (2008).

[7] Z. J. Jiang, Q. G. Li and H. L. Zhang, *J*. *Phys*. *G*: *Nucl*. *Part*. *Phys*. **40** 025101 (2013).

[8] C. Gale, S. Jeon and B. Schenke, *Intern. J. Mod. Phys.* A **28** 1340011 (2013).

[9] E. K. G. Sarkisyan and A. S. Sakharov, *Eur*. *Phys*. *J.* C **70** 533 (2010).

[10] H. Song, Steffen A. Bass, U. Heinz, T. Hirano and C. Shen, *Phys*. *Rev*. *Lett*. **106** 192301 (2011).

[11] R. Rvblewski and W. Florkowski, *Phys*. *Rev*. C **85** 064901 (2012).

[12] L. D. Landau, *Izv*. *Akad*. *Nauk SSSR* **17** 51 (1953) (in Russian).

[13] I. M. Khalatnikov, *J*. *Exp*. *Theor*. *Phys*. **27** 529 (1954) (in Russian).

[14] S. Z. Belenkij and L. D. Landau, *Nuovo Cimento Suppl*. **3** 15 (1956).

[15] M. Murray (for BRAHMS Collaboration), *J*. *Phys*. *G*: *Nucl*. *Part*. *Phys*. **30** S667 (2004).

[16] M. Murray (for BRAHMS Collaboration), *J*. *Phys*. *G*: *Nucl*. *Part*. *Phys*. **35** 044015 (2008).

[17] I. G. Bearden *et al.* (BRAHMS Collaboration), *Phys*. *Rev*. *Lett*. **94** 162301 (2005).

[18] R. C. Hwa, *Phys*. *Rev*. D **10** 2260 (1974).

[19] J. D. Bjorken, *Phys*. *Rev*. D **27** 140 (1983).

[20] G. Beuf, R. Peschanski and E. N. Saridakis, *Phys*. *Rev*. C **78** 064909 (2008).

[21] T. Csörgő, N. I. Nagy and M. Csanád, *Phys*. *Lett*. B **663** 306 (2008).

[22] M. I. Nagy, T. Csörgő and M. Csanád, *Phys*. *Rev*. C **77** 024908 (2008).

[23] M. Csanád, M. I. Nagy and T. Csörgő, *Eur*. *Phys*. *J. ST* **155** 19 (2008).

[24] A. Bialas, R. A. Janik and R. Peschanski, *Phys*. *Rev*. C **76** 054901 (2007).

[25] M. S. Borshch and V. I. Zhdanov, *SIGMA* **3** 116 (2007).

[26] M. Csanád, M. I. Nagy and S. Lökös, *Eur*. *Phys*. *J*. A **48** 173 (2012).

[27] B. Alver *et al.* (PHOBOS Collaboration), *Phys. Rev.* C **83** 024913 (2011).

[28] E. Abbas *et al.* (ALICE Collaboration), *Phys. Lett.* B **726** 610 (2013).

[29] A. Adare *et al.* (PHENIX Collaboration), *Phys. Rev. Lett.* **98** 162301 (2007).

[30] N. Armesto, N. Borghini and S. Jeon *et al.*, *J. Phys. G: Nucl. Part. Phys.* **35** 054001 (2008).

[31] T. Mizoguchi, H. Miyazawa and M. Biyajima, *Eur. Phys. J.* A **40** 99 (2009).

[32] S. Borsányi, G. Endrődi, Z. Fodor, A. Jakovác, S. D. Katz, S. Krieg, C. Ratti and K. K. Szabó *JHEP* **77** 1 (2010).

[33] S. Amai, H. Fukuda, C. Iso and M. Sato, *Prog. Theor. Phys.* **17** 241 (1957).

[34] T. B. Li, *The Mathematical Processing of Experiments*, Science Press: Beijing, p.42 (1980) (in Chinese).

[35] J. Voit, *The statistical mechanics of financial markets*, Springer: Berlin, p.123 (2005).

[36] Z. J. Jiang, Y. F. Sun and Q. G. Li, *Int. J. Mod. Phys.* E **21** 1250002 (2012).

[37] B. B. Back *et al.* (PHOBOS Collaboration), *Nucl. Phys.* A **757** 28 (2005).

[38] K. Aamodt *et al.* (ALICE Collaboration), *Phys. Rev. Lett.* **106** 032301 (2011).

[39] Z. J. Jiang, *Acta Physica Sinica* **56** 5191 (2007).

[40] C. Y. Wong, *Introduction to High Energy Heavy Ion Collisions*, Press of Harbin Technology University: Harbin, p. 23 (2002) (in Chinese); English edition: World Scientific: Singapore, p. 25 (1994).

[41] I. G. Bearden *et al.* (BRAHMAS Collaboration), *Phys. Rev. Lett.* **93** 102301 (2004).

**Table and figure captions**

**Table 1**

The mean numbers of total participants $\bar{N}_{Part}$ and the central positions $y_0$ of Gaussian rapidity distribution in different centrality Au-Au and Cu-Cu collisions at $\sqrt{s_{NN}} = 200$ GeV. The numbers with and without errors are respectively the results given by PHOBOS Collaboration at BNL-RHIC [27] and Eq. (11).

**Table 2**

The mean numbers of total participants $\bar{N}_{Part}$ and the central positions $y_0$ of Gaussian rapidity distribution in different centrality Pb-Pb collisions at $\sqrt{s_{NN}} = 2.76$ TeV. The numbers with and without errors are respectively the results given by ALICE Collaboration at CERN-LHC [28] and Eq. (11).

**Figure 1**

The rapidity distributions of specified charged particles in central Au-Au collisions at $\sqrt{s_{NN}} = 200$ GeV. The scattered symbols are the experimental measurements [15-17]. The solid curves are the results from the evolution-dominated hydrodynamics of Eq. (9).

**Figure 2**

The pseudorapidity distributions of the charged particles produced in different centrality Au-Au collisions at $\sqrt{s_{NN}} = 200$ GeV. The solid dots are the experimental measurements [27]. The dashed curves are the results from evolution-dominated hydrodynamics of Eq. (9). The dotted curves are the results from leading particles of Eq. (10). The solid curves are the sums of dashed and dotted ones.

**Figure 3**

The pseudorapidity distributions of the charged particles produced in different centrality Cu-Cu collisions at $\sqrt{s_{NN}} = 200$ GeV. The solid dots are the experimental measurements [27]. The dashed curves are the results from evolution-dominated hydrodynamics of Eq. (9). The dotted curves are the results from leading particles of Eq. (10). The solid curves are the sums of dashed and dotted ones.

**Figure 4**

The pseudorapidity distributions of the charged particles produced in different centrality Pb-Pb collisions at $\sqrt{s_{NN}} = 2.76$ TeV. The solid dots are the experimental measurements [28]. The dashed curves are the results from evolution-dominated hydrodynamics of Eq. (9). The dotted curves are the results from leading particles of Eq. (10). The solid curves are the sums of dashed and dotted ones.

**Table 1**

| Centrality Cut (%) | 0-3 | 3-6 | 6-10 | 10-15 | 15-20 | 20-25 | 25-30 | 30-35 | 35-40 |
|---|---|---|---|---|---|---|---|---|---|
| $\bar{N}_{Part}$(Au-Au) | 359.44 | 324.50 | 288.74 | 248.96 | 210.98 | 178.24 | 149.78 | 124.92 | 103.22 |
| | 361±11 | 331±10 | 297±9 | 255±8 | 215±7 | 180±7 | 150±6 | 124±6 | 101±6 |
| $\bar{N}_{Part}$(Cu-Cu) | 109.92 | 99.76 | 89.00 | 76.70 | 64.74 | 54.40 | 45.40 | 37.62 | 30.88 |
| | 108±4 | 101±3 | 91±3 | 79±3 | 67±3 | 57±3 | 48±3 | 40±3 | 33±3 |
| $y_0$(Au-Au) | 2.63 | 2.67 | 2.70 | 2.72 | 2.78 | 2.81 | 2.96 | 2.97 | 3.05 |
| $y_0$(Cu-Cu) | 2.75 | 2.78 | 2.80 | 2.93 | 2.94 | 2.95 | 2.96 | 2.97 | 3.05 |

**Table 2**

| Centrality Cut (%) | 0-5 | 5-10 | 10-20 | 20-30 |
|---|---|---|---|---|
| $\bar{N}_{Part}$(Pb-Pb) | 381.56 | 327.70 | 261.90 | 189.78 |
| | 383±3 | 330±5 | 261±4 | 186±4 |
| $y_0$(Pb-Pb) | 3.38 | 3.41 | 3.44 | 3.48 |

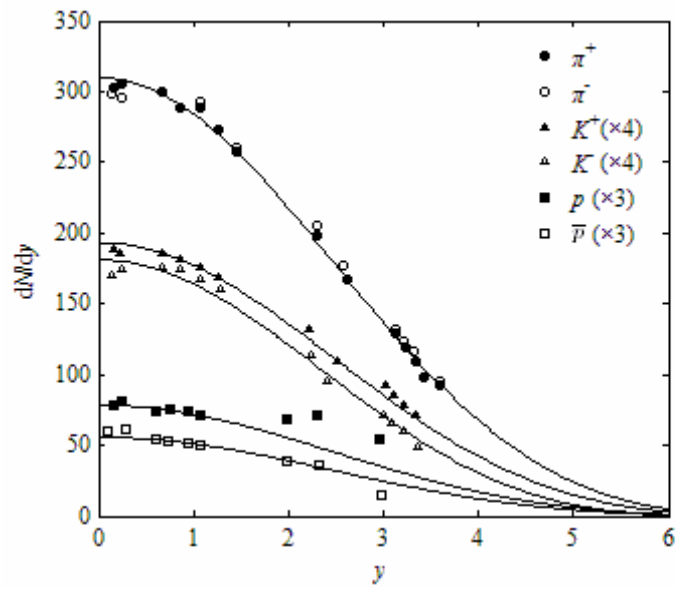

**Figure 1**

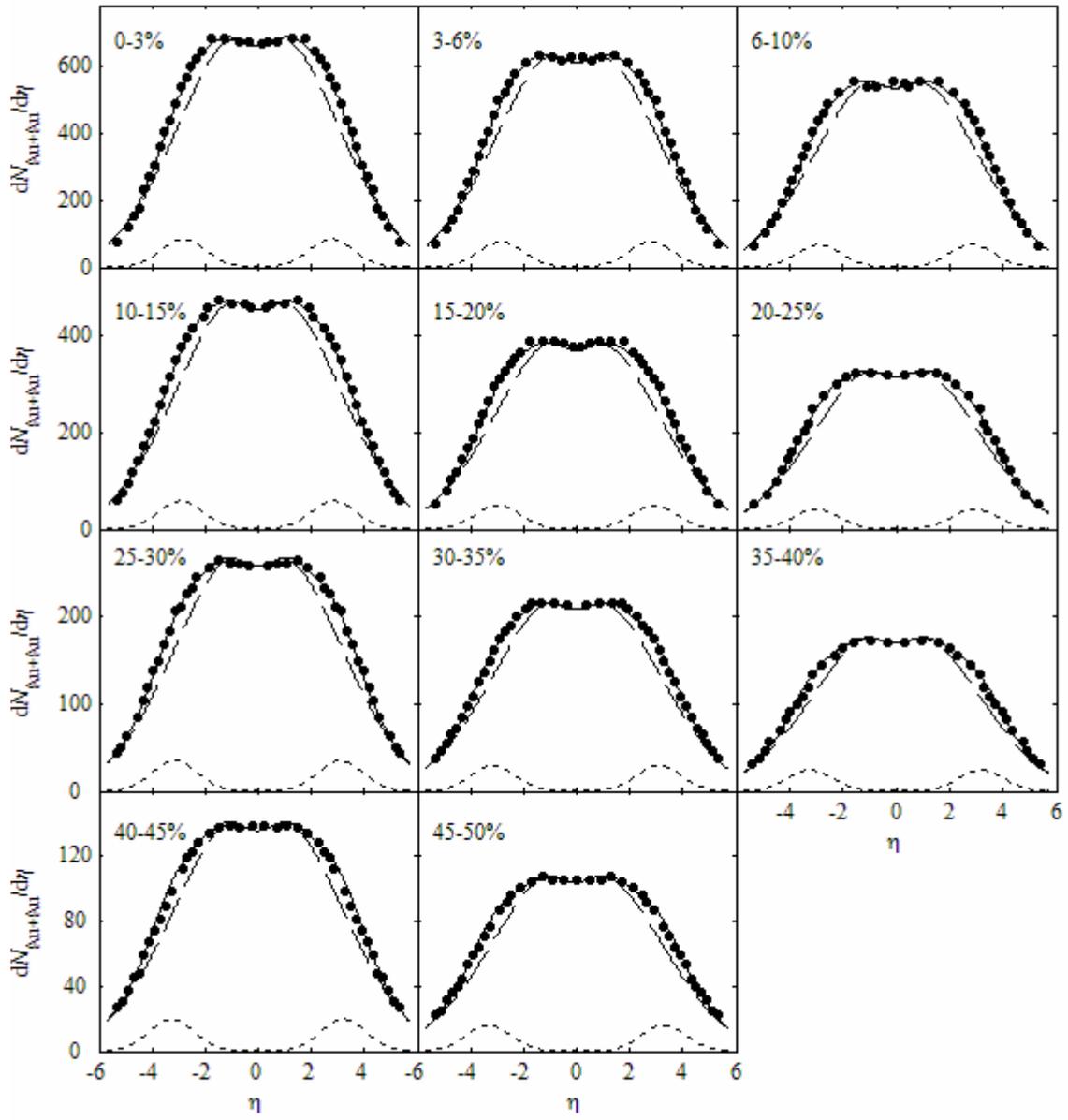

**Figure 2**

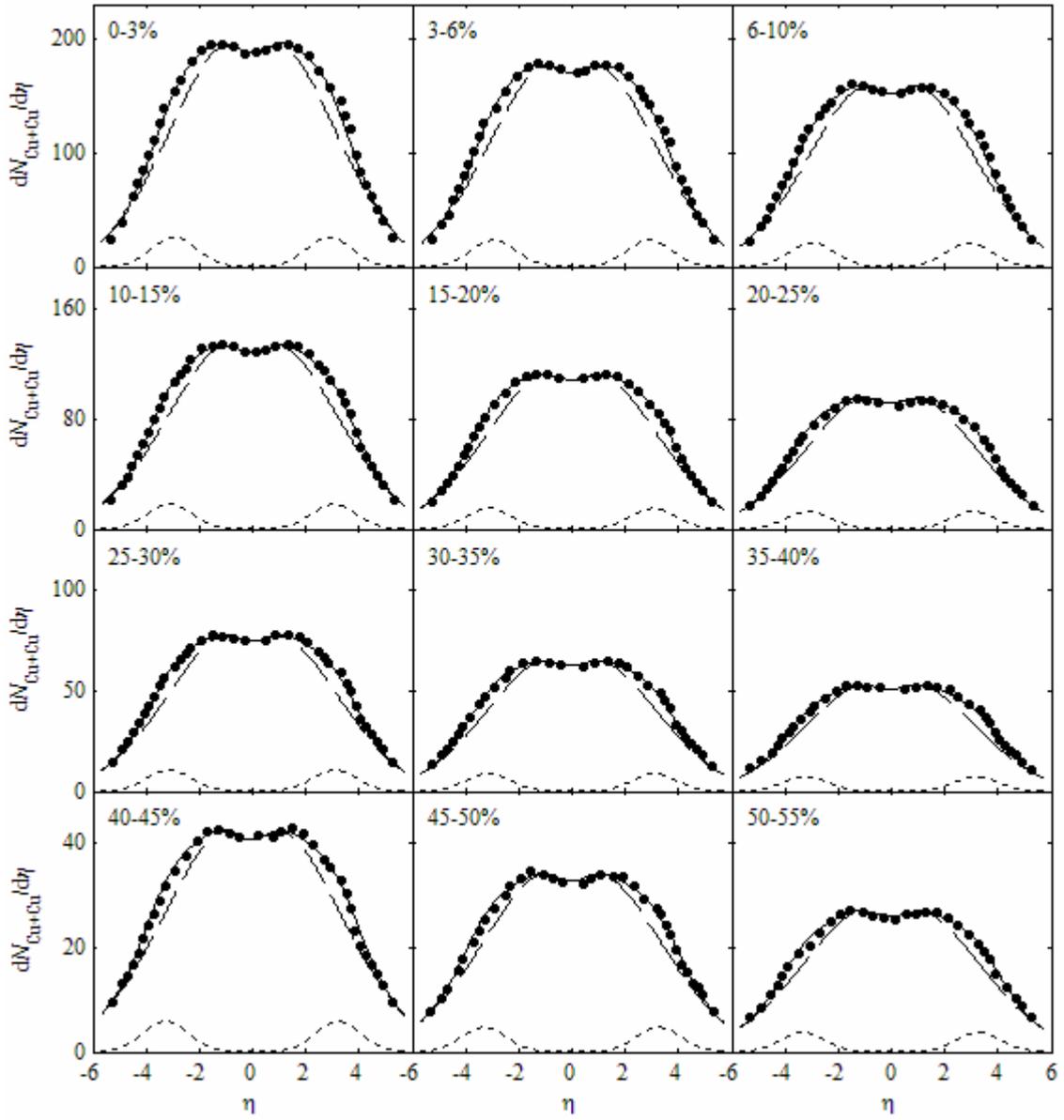

**Figure 3**

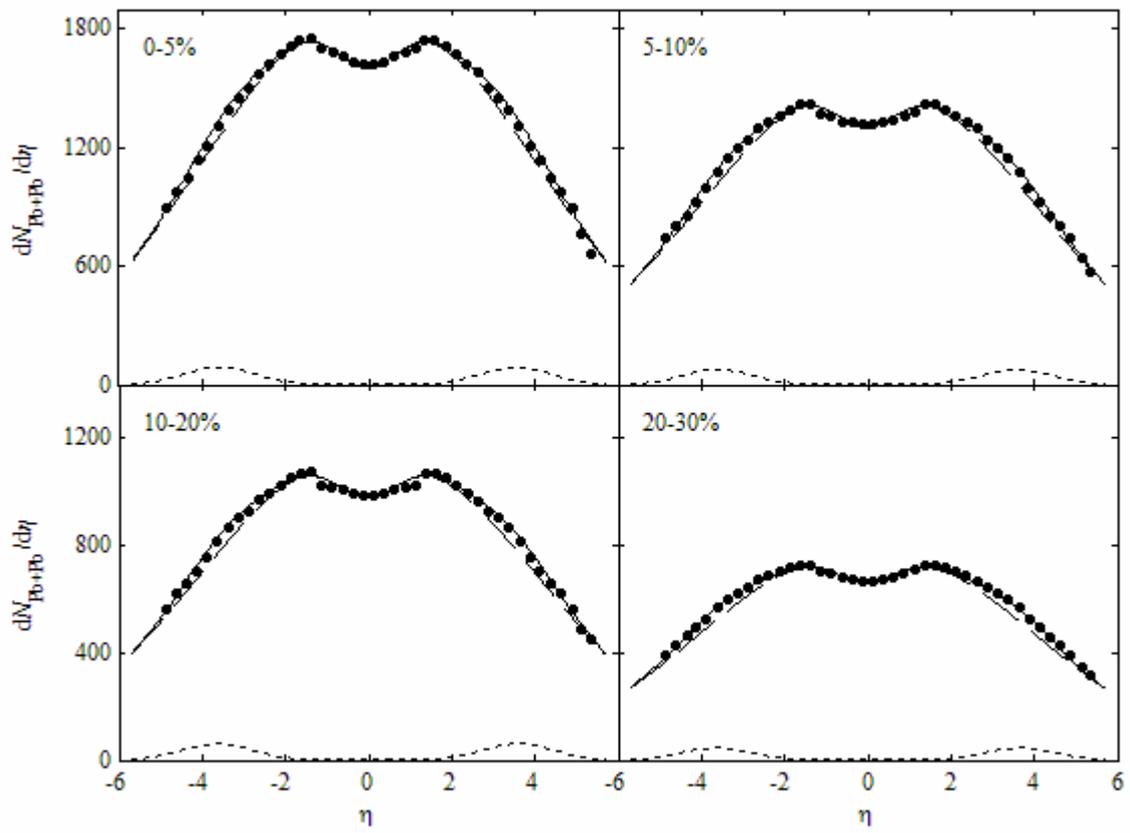

**Figure 4**